\documentclass{article}
\usepackage{spconf,amsmath,amssymb,graphicx,hyperref,booktabs,multirow}
\usepackage{url}
\usepackage{orcidlink}

\usepackage{eso-pic}
\usepackage{ragged2e}

\newcommand{\IEEEcopyrightnotice}{%
\textcopyright\ 2026 IEEE. Personal use of this material is permitted. Permission from IEEE must be obtained for all other uses, in any current or future media, including reprinting/republishing this material for advertising or promotional purposes, creating new collective works, for resale or redistribution to servers or lists, or reuse of any copyrighted component of this work in other works.
}

\AddToShipoutPictureFG*{%
  \AtPageLowerLeft{%
    \raisebox{1.3cm}[0pt][0pt]{%
      \hspace*{\dimexpr 1in+\hoffset+\oddsidemargin\relax}%
      \parbox{\textwidth}{%
        \fontsize{7.7}{8.9}\selectfont
        \justifying
        \setlength{\parindent}{0pt}%
        \emergencystretch=1em
        \noindent
        \IEEEcopyrightnotice
      }%
    }%
  }%
}

\title{Adaptive Depth and Expert Refinement for Efficient Speech Enhancement}
\name{
    Xikun Lu\orcidlink{0000-0003-0156-8805}$^{1}$ \qquad
    Yujian Ma\orcidlink{0009-0006-1652-5903}$^{1}$ \qquad
    Yunda Chen\orcidlink{0000-0003-4470-7262}$^{2}$ \qquad
    Xianquan Jiang\orcidlink{0009-0009-4360-4836}$^{3}$ \qquad
    Jinqiu Sang\orcidlink{0000-0002-4368-8787}$^{4\ast}$\thanks{${\ast}$ Corresponding author: Jinqiu Sang (jqsang@mail.ecnu.edu.cn).}
}

\address{%
  $^{1}$ Shanghai Institute of Artificial Intelligence for Education, East China Normal University, China\\
  $^{2}$ Guangdong Key Laboratory of Intelligent Information Processing, Shenzhen University, China\\
  $^{3}$ Boin Hearing Technology (Shanghai) Co., LTD, China\\
  $^{4}$ School of Computer Science and Technology, East China Normal University, China
}
\begin{document}
\ninept
\maketitle
\begin{abstract}

%
Most neural speech enhancement systems use a fixed processing depth for all inputs, which can introduce unnecessary computation when fewer refinement steps are sufficient. We propose \emph{Adaptive Depth and Expert Refinement} (ADER), a parameter-shared progressive enhancement framework with input-dependent computation. ADER combines an \emph{Adaptive Depth Controller} (ADC) for hard early termination with a \emph{Conditional Expert Router} (CER) that selects one lightweight residual adapter at each executed refinement iteration. We further introduce \emph{Exit-aware Intermediate Supervision} (EIS) to directly optimize candidate intermediate outputs for early exit. On VCTK-DEMAND, ADER reduces the parameter count and average computation of MP-SENet by 70.4\% and 51.3\%, respectively, while achieving a WB-PESQ of 3.37. Overall, ADER enables input-dependent refinement and reduces redundant computation during inference.
\end{abstract}

\section{Introduction}
\label{sec:intro}

Deep learning has driven substantial progress in monaural speech enhancement (SE), including complex-valued spectral modeling, phase-aware estimation, and time--frequency (TF) sequence modeling \cite{hu2020dccrn,yin2020phasen,fu2021metricganplus,
AbdulatifCY24,lu2025mpsenet}. Recent studies have further examined the effects of network architecture, model size, and computational budget on SE performance \cite{zhang2024scalability}. As SE systems are increasingly deployed under real-time and resource constraints, computational efficiency has become an important design consideration. A central problem is therefore to reduce inference cost while retaining effective TF modeling for SE.

Existing studies have addressed this problem by designing more efficient SE architectures. FullSubNet combines full-band and sub-band modeling for real-time enhancement \cite{hao2021fullsubnet}, while DeepFilterNet uses deep filtering and grouped operations for low-complexity full-band processing \cite{schroter2022deepfilternet}. BLOOM-Net further introduces blockwise optimization to produce scalable intermediate enhancement outputs \cite{kim2022bloom}. A different approach is to reduce parameter redundancy through block reuse. Kim \emph{et al.} proposed a progressive SE framework that repeatedly applies a shared processing block instead of stacking independently parameterized blocks \cite{kim2025stack}. This design enables progressive refinement with substantially fewer independent parameters. However, the repetition depth is predefined and remains fixed during inference. Consequently, all utterances undergo the same number of refinement iterations, even when fewer iterations may be sufficient for some inputs.

Dynamic computation provides a direct way to adapt the number of executed refinement steps. Adaptive Computation Time introduced input-dependent computational depth in recurrent networks \cite{graves2016act}, and related ideas have subsequently been applied to speech processing. Li \emph{et al.} introduced early exit into progressive SE using the difference between consecutive enhancement outputs as a stopping criterion \cite{li2021earlyexit}. Dynamic nsNet2 incorporates multiple exits into a noise suppression network to vary inference cost across inputs \cite{miccini2023dynamic}. Early-exit Transformers have also been studied for continuous speech separation \cite{chen2021earlyexit}. More recently, Olsen \emph{et al.} introduced probabilistic stopping criteria for speech separation and enhancement \cite{olsen2026knowing}. These studies demonstrate that processing depth can be adapted to individual inputs. For a block-reuse architecture, however, adaptive depth alone does not address the limited flexibility of full parameter sharing. As refinement proceeds, the latent representation evolves, whereas the same shared block is applied at every iteration. Sparse expert routing can complement the shared block with lightweight state-dependent adaptation \cite{shazeer2017moe,fedus2022switch}. This suggests that block-reuse models should adapt the refinement depth to each input and the refinement operation to the current state.

\begin{figure*}
    \centering
    \includegraphics[width=1.0\linewidth]{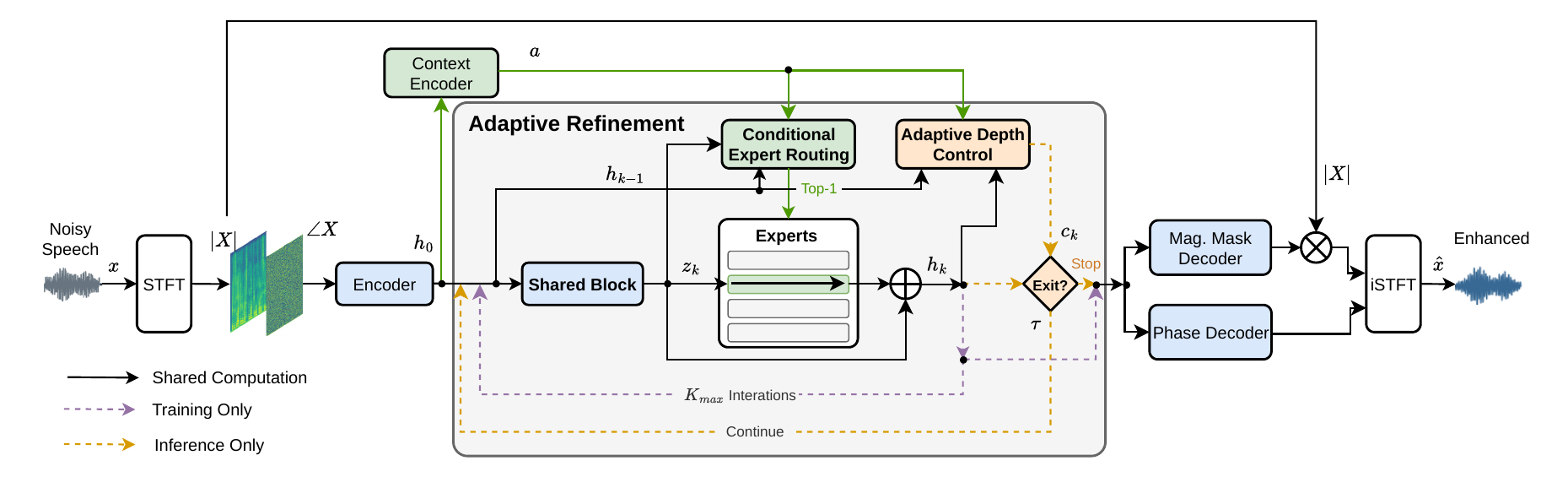}
    \caption{Overview of the proposed ADER framework. A shared refinement block is repeatedly applied to the encoded TF representation. At each iteration, the CER selects one residual expert adapter, and the ADC determines whether another refinement iteration is required.}
    \label{fig:overview}
\end{figure*}


Inspired by the roles of cellular proliferation and differentiation, which respectively correspond to continuing refinement when needed and applying specialized processing to the current state, we propose \emph{Adaptive Depth and Expert Refinement} (ADER)\footnote{ Our source codes are available online: \url{https://github.com/Luxikun669/ADER}}, which combines a shared refinement backbone with adaptive depth control and conditional expert routing. Specifically, our contributions are threefold.  
1) We introduce an \emph{Adaptive Depth Controller} (ADC) that determines whether another refinement iteration should be executed, allowing the number of repeated refinement steps to vary across utterances. 2) We introduce a \emph{Conditional Expert Router} (CER) that selects one lightweight residual adapter at each executed iteration through straight-through Top-1 routing. This provides state-dependent adaptation while keeping the main refinement block shared. 3) We introduce \emph{Exit-aware Intermediate Supervision} (EIS), which applies the complete enhancement objective to a randomly sampled intermediate iteration during training, making intermediate outputs more reliable for hard early exit. Experiments on VCTK-DEMAND evaluate the resulting adaptive computation, refinement-depth behavior, and iteration-dependent expert routing.

\section{Proposed method}
\label{sec:method}

Figure~\ref{fig:overview} presents the overall architecture of the proposed ADER framework. ADER is built on a parameter-shared progressive refinement backbone, with the CER and ADC controlling the expert transformation and refinement depth, respectively. 

\subsection{Shared Progressive Refinement}
\label{sec:shared}

Let $x\in\mathbb{R}^{L}$ denote a noisy speech waveform with $L$ samples. Following the magnitude--phase front end of MP-SENet \cite{LuAL23}, its STFT magnitude and phase are processed by the encoder $E(\cdot)$ to obtain an initial TF representation $h_0\in\mathbb{R}^{C\times T\times F}$, where $C$, $T$, and $F$ denote the feature-channel, time-frame, and frequency-bin dimensions, respectively.

The encoded representation $h_0$ is used in two ways. First, it serves as the initial state of the progressive refinement process. Second, a lightweight context encoder extracts $a=E_{\rm ctx}(h_0)$, which provides conditioning information to both CER and ADC throughout the refinement process. The context encoder $E_{\rm ctx}$ consists of a $1\times1$ convolution, instance normalization, and PReLU.

ADER performs progressive refinement using a single TS-Transformer block adopted from MP-SENet~\cite{LuAL23} as the shared refinement block $C_\theta$. The block models temporal and spectral dependencies in the TF representation and is recursively reused across refinement iterations. At iteration $k$,
\begin{equation}
    z_k=C_\theta(h_{k-1}),
    \qquad k=1,\ldots,K_{\max},
    \label{eq:shared}
\end{equation}
where $h_{k-1}$ is the current refinement state, $z_k$ is the shared-block output before expert adaptation, and $K_{\max}$ denotes the maximum refinement depth. For $k=1$, $h_{k-1}=h_0$; subsequent iterations use the refined representation from the preceding iteration. Since the parameters of $C_\theta$ are reused across iterations, increasing the refinement depth does not introduce additional copies of the refinement block.

\subsection{Conditional Expert Routing}
\label{sec:cer}

Although the backbone parameters are shared, the representation evolves after each refinement iteration. ADER therefore introduces a lightweight conditional expert branch to adapt the shared-block output according to the current refinement state.

At refinement iteration $k$, the CER forms its routing descriptor from $h_{k-1}$, $z_k$, and the acoustic context $a$. After global average pooling,
\begin{equation}
q_k^{\rm CER}
=
\left[
P(h_{k-1}),
P(z_k),
P(a)
\right],
\label{eq:cer_descriptor}
\end{equation}
where $P(\cdot)$ denotes global average pooling and $[\cdot]$ denotes feature concatenation. The absolute-difference term reflects the change introduced by the shared refinement block, while $a$ provides utterance-level context throughout the refinement process.

The router $R_{\rm CER}$ maps $q_k^{\rm CER}$ to an expert probability distribution
\begin{equation}
\pi_k
=
\operatorname{softmax}
\left(
\frac{R_{\rm CER}(q_k^{\rm CER})}{T_r}
\right),
\label{eq:cer_prob}
\end{equation}
where $T_r$ denotes the routing temperature. We use $M$ lightweight expert adapters $\{A_m\}_{m=1}^{M}$. Each adapter contains a bottleneck $1\times1$ projection, a depthwise $3\times3$ convolution, and a $1\times1$ output projection, together with a learnable residual scale. This design keeps the expert branch small relative to the shared refinement block.

During training, straight-through Gumbel Top-1 routing produces a one-hot routing vector $r_k\in\{0,1\}^{M}$. The output of the $k$-th refinement iteration is
\begin{equation}
    h_k
    =
    z_k+
    \sum_{m=1}^{M}r_{k,m}A_m(z_k).
    \label{eq:cer_refine}
\end{equation}

Thus, the shared block is executed at every refinement iteration, while only one expert adapter contributes to the forward output. During inference, the selected expert is $m_k^\star=\arg\max_m\pi_{k,m}$, and only $A_{m_k^\star}$ is evaluated.

To avoid routing collapse, we apply a Switch-style load-balancing loss independently at each refinement iteration. For expert $m$, let $p_{k,m}$ denote its mean soft routing probability and $\ell_{k,m}$ its hard selection frequency within the current batch. The balancing term is
\begin{equation}
\mathcal{L}_{\rm bal}
=
\frac{M}{K_{\max}}
\sum_{k=1}^{K_{\max}}
\sum_{m=1}^{M}
\operatorname{sg}(\ell_{k,m})p_{k,m},
\label{eq:balance}
\end{equation}
where $\operatorname{sg}(\cdot)$ denotes stop-gradient. Computing this term separately for each iteration encourages routing diversity without requiring different copies of the main refinement block.

\subsection{Adaptive Depth Control}
\label{sec:adc}

Conditional expert routing changes the residual transformation used at each refinement iteration, but it does not determine how many iterations should be executed. We therefore introduce an ADC to decide whether the current representation requires further refinement.

After each refinement iteration, the ADC determines whether further refinement is required using $h_{k-1}$, $h_k$, and the acoustic context $a$. Its input descriptor is
\begin{align}
q_k^{\rm ADC}
&=
\left[
P(h_{k-1}),
P(h_k),
P(a)
\right],
\end{align}
\begin{align}
c_k
&=
\sigma\!\left(R_{\rm ADC}(q_k^{\rm ADC})\right),
\label{eq:adc}
\end{align}
where $R_{\rm ADC}$ denotes the depth controller, $\sigma(\cdot)$ is the sigmoid function, and $c_k\in[0,1]$ is the probability of continuing to another refinement iteration.

ADER executes at least $K_{\min}$ and at most $K_{\max}$ refinement iterations, where $K_{\min}$ and $K_{\max}$ denote the minimum and maximum refinement depths, respectively. Once $k\ge K_{\min}$, inference terminates when $c_k<\tau$, where $\tau$ is the continuation threshold; otherwise, refinement proceeds to iteration $k+1$. The resulting input-dependent refinement depth is therefore $N(x)\in\{K_{\min},\ldots,K_{\max}\}$.

The ADC requires a training target indicating whether one more refinement iteration is useful. We derive this target from the reconstruction quality along the full refinement trajectory. For the $k$-th iteration, we use
\begin{equation}
Q_k
=
0.9\mathcal{L}_{\rm mag}^{(k)}
+
0.3\mathcal{L}_{\rm pha}^{(k)}
+
0.1\mathcal{L}_{\rm com}^{(k)},
\label{eq:quality_proxy}
\end{equation}
where $\mathcal{L}_{\rm mag}^{(k)}$, $\mathcal{L}_{\rm pha}^{(k)}$, and $\mathcal{L}_{\rm com}^{(k)}$ denote the magnitude, phase, and complex-spectrum reconstruction losses obtained from the output at iteration $k$.

Let $Q_{\rm best}=\min_{j\ge K_{\min}}Q_j$ denote the best reconstruction quality among the valid exit iterations. Rather than forcing every sample to reach the globally best iteration, we select the earliest iteration whose quality is sufficiently close to this minimum:
\begin{equation}
N^\star
=
\min
\left\{
k\ge K_{\min}
\,\middle|\,
Q_k\le(1+\delta)Q_{\rm best}
\right\},
\label{eq:oracle_depth}
\end{equation}
where $\delta$ controls the allowed reconstruction difference from the best iteration. The continuation target is then $t_k=\mathbb{I}[k<N^\star]$, where $\mathbb{I}[\cdot]$ is the indicator function.

Only decisions that would be reached before the target exit are included when training the ADC. For sample $i$, we use the reachability weight $w_{i,k}^{\rm reach} =\mathbb{I}[K_{\min}\le k\le N_i^\star]$.
The depth-control objective is
\begin{equation}
\mathcal{L}_{\rm ADC}
=
\frac{1}{|\mathcal K|}
\sum_{k\in\mathcal K}
\frac{
\sum_i
w_{i,k}^{\rm reach}
\,\mathrm{BCE}(c_{i,k},t_{i,k})
}{
\sum_i w_{i,k}^{\rm reach}+\epsilon
},
\label{eq:adc_loss}
\end{equation}
where $\mathcal K=\{K_{\min},\ldots,K_{\max}-1\}$ is the set of actionable decisions, $\mathrm{BCE}(\cdot)$ denotes binary cross entropy, and $\epsilon$ is a small constant for numerical stability.

\subsection{Exit-Aware Intermediate Supervision and Training}
\label{sec:eis}

Reliable intermediate outputs are required for hard early exit. However, under full-depth training, task supervision is applied primarily to the final refinement output, while intermediate states are optimized mainly through subsequent refinement. This creates a mismatch between training and adaptive-depth inference. 

To reduce this mismatch, we introduce EIS. During training, ADER always unfolds all $K_{\max}$ refinement iterations and uniformly samples one intermediate iteration $d\in\{K_{\min},\ldots,K_{\max}-1\}$. The corresponding representation $h_d$ is decoded by the same magnitude-mask and phase decoders used for the final iteration and is directly supervised by the complete enhancement objective.

For compact notation, the enhancement objective at iteration $k$ is defined as
\begin{align}
\mathcal{L}_{\rm enh}^{(k)}
={}&
0.9\mathcal{L}_{\rm mag}^{(k)}
+
0.3\mathcal{L}_{\rm pha}^{(k)}
+
0.1\mathcal{L}_{\rm com}^{(k)}
\nonumber\\
&+
0.1\mathcal{L}_{\rm stft}^{(k)}
+
0.05\mathcal{L}_{\rm metric}^{(k)}
+
0.2\mathcal{L}_{\rm time}^{(k)},
\label{eq:enh_loss}
\end{align}
where $\mathcal{L}_{\rm stft}$, $\mathcal{L}_{\rm metric}$, and $\mathcal{L}_{\rm time}$ denote the multi-resolution STFT, metric-based perceptual, and time-domain reconstruction losses, respectively. The final and intermediate supervision terms are $\mathcal{L}_{\rm MP}=\mathcal{L}_{\rm enh}^{(K_{\max})}$ and $\mathcal{L}_{\rm EIS}=\mathcal{L}_{\rm enh}^{(d)}$, respectively. The deepest iteration is excluded from EIS sampling because it is already optimized by $\mathcal{L}_{\rm MP}$.

The overall training objective is
\begin{equation}
\mathcal{L}
=
\mathcal{L}_{\rm MP}
+
\lambda_{\rm EIS}\mathcal{L}_{\rm EIS}
+
\lambda_{\rm bal}\mathcal{L}_{\rm bal}
+
\lambda_{\rm ADC}\mathcal{L}_{\rm ADC},
\label{eq:total}
\end{equation}
where $\lambda_{\rm EIS}=0.1$,
$\lambda_{\rm bal}=2.5\times10^{-4}$, and
$\lambda_{\rm ADC}=0.05$.

Training is conducted in two stages. An initial 40-epoch warm-up stage optimizes the enhancement, intermediate-supervision, and routing-balancing objectives without $\mathcal{L}_{\rm ADC}$. The complete objective in Eq.~(\ref{eq:total}) is subsequently used for joint optimization. All $K_{\max}=5$ refinement iterations are unrolled during training, whereas hard early exit is enabled only at inference.

\begin{table*}[t]
\centering
\caption{Overall comparison on VCTK-DEMAND. MP-SENet* denotes the reduced-depth MP-SENet with $D=2$. $D$ denotes the dense encoder (and decoder) depth, while $\bar N$ denotes the mean number of executed refinement iterations for block-reuse models. \textbf{BOLD} indicates the best score in each metric.}
\label{tab:overall}
\setlength{\tabcolsep}{4.0pt}
\begin{tabular}{lcccccccccc}
\toprule
Method & $D$ & $\bar N$ & Param. (M) & GMACs/s &
WB-PESQ & STOI (\%) & SSNR & CSIG & CBAK & COVL \\
\midrule

MP-SENet~\cite{LuAL23}
& 4 & 4.00 & 2.26 & 42.59
& \textbf{3.50} & \textbf{96.15} & \textbf{10.64}
& \textbf{4.73} & \textbf{3.95} & \textbf{4.22} \\

MP-SENet*~\cite{LuAL23}
& 2 & 4.00
& 1.75\,{\tiny ($\downarrow$22.6\%)}
& 31.43\,{\tiny ($\downarrow$26.2\%)}
& 3.47\,{\tiny ($\downarrow$0.9\%)} & 96.05 & 10.60
& 4.71 & 3.93 & 4.20 \\

Fixed Shared-Block
& 2 & 3.00
& \textbf{0.65}\,{\tiny ($\downarrow$71.2\%)}
& 25.02\,{\tiny ($\downarrow$41.3\%)}
& 3.46\,{\tiny ($\downarrow$1.1\%)} & 95.85 & 10.52
& 4.70 & 3.92 & 4.19 \\

\textbf{ADER} ($\tau=0.20$)
& 2 & 2.33
& 0.67\,{\tiny ($\downarrow$70.4\%)}
& 21.99\,{\tiny ($\downarrow$48.4\%)}
& 3.39\,{\tiny ($\downarrow$3.1\%)} & 95.71 & 10.34
& 4.69 & 3.88 & 4.14 \\

\textbf{ADER} ($\tau=0.50$)
& 2 & \textbf{2.15}
& 0.67\,{\tiny ($\downarrow$70.4\%)}
& \textbf{20.73}\,{\tiny ($\downarrow$51.3\%)}
& 3.37\,{\tiny ($\downarrow$3.7\%)} & 95.67 & 10.30
& 4.68 & 3.87 & 4.13 \\

\bottomrule
\end{tabular}
\end{table*}

\section{Experiments and Results}
\label{sec:exp}

\subsection{Experimental Setup}

We evaluate the proposed ADER framework on the standard VCTK-DEMAND corpus \cite{valentini2017noisy,10.1121/1.4799597}, using its 824-utterance test set. All audio is resampled to 16~kHz. STFT analysis uses a 400-sample window with a 100-sample hop. The dense channel dimension is set to 64 and the dense encoder depth to $D=2$. For ADER, the minimum and maximum refinement depths are set to $K_{\min}=2$ and $K_{\max}=5$, respectively. We use four expert adapters, each with a bottleneck channel dimension of $C/4$.

All models are trained for 100 epochs using AdamW with a learning rate of $5\times10^{-4}$, $\beta_1=0.8$, $\beta_2=0.99$, with a global batch size of 4. The CER uses straight-through Gumbel Softmax with hard Top-1 selection during training and deterministic Top-1 routing during inference. The ADC objective is activated after an initial 40-epoch warm-up stage.


We report wideband PESQ (WB-PESQ) \cite{rix2001pesq}, STOI \cite{taal2010stoi,5713237}, segmental SNR (SSNR), and the composite measures CSIG (signal distortion), CBAK (background noise intrusiveness), and COVL (overall quality) \cite{hu2008objective}. Model complexity is measured by the number of parameters and the mean multiply--accumulate operations per second of input audio (MACs/s). For ADER, the MAC count is computed from the refinement iterations actually executed for each utterance and then averaged over the test set.

\subsection{Overall Comparison}

Table~\ref{tab:overall} compares ADER with the original MP-SENet, its reduced-depth MP-SENet*, and the fixed shared-block baseline. ADER contains 0.67~M parameters, reducing the parameter count of MP-SENet by 70.4\%. With $\tau=0.20$, ADER executes 2.33 refinement iterations on average and requires 21.99~GMACs/s, corresponding to a 48.4\% reduction in computation relative to MP-SENet. Increasing the threshold to $\tau=0.50$ further reduces the mean refinement depth to 2.15 and the average computation to 20.73~GMACs/s, a 51.3\% reduction relative to MP-SENet.

Compared with the fixed shared-block model at $\bar N=3$, ADER trades a moderate decrease in enhancement quality for lower average computation. At $\tau=0.20$, the computational cost decreases from 25.02 to 21.99~GMACs/s, a reduction of 12.1\%, while WB-PESQ decreases by about 2.0\% from 3.46 to 3.39. With $\tau=0.50$, the computation is further reduced by 17.1\% to 20.73~GMACs/s, with a corresponding 2.6\% decrease in WB-PESQ. Unlike the fixed-depth model, ADER selects the executed refinement depth for each utterance, allowing the average computational cost to be adjusted through the continuation threshold.

\subsection{Component Analysis}

\begin{table}[t]
\centering
\caption{Component analysis of ADC, CER, and EIS. Adaptive variants use $\tau=0.50$. \textbf{BOLD} indicates the best score in each metric.}
\label{tab:ablation}
\scriptsize
\setlength{\tabcolsep}{4.2pt}
\begin{tabular}{cccccccc}
\toprule
ADC & CER & EIS  & $\bar N$ & Param. (M) & GMACs/s &
WB-PESQ & STOI (\%)  \\
\midrule

-- & -- & --
&  3.00 & \textbf{0.65}  & 25.02 
& \textbf{3.46} & 95.85  \\

\checkmark & -- & --
& 2.92 & 0.66 & 25.95 
& 3.32 & 95.74  \\

-- & \checkmark & --
&  3.00 & 0.67 & 25.19 
& 3.43 & \textbf{95.81}  \\

\checkmark & \checkmark & --
& 2.93 &  0.67 & 26.71 
& 3.40 & 95.77  \\

\checkmark & \checkmark & \checkmark
&  \textbf{2.15} & 0.67 & \textbf{20.73}
& 3.37 & 95.67  \\

\bottomrule
\end{tabular}
\end{table}

Table~\ref{tab:ablation} further examines the contributions of ADC, CER, and EIS. Introducing ADC alone reduces WB-PESQ from 3.46 to 3.32, indicating that direct early termination makes intermediate representations a major source of quality degradation. CER alone causes a smaller reduction, yielding a WB-PESQ of 3.43. When CER is combined with ADC, WB-PESQ increases from 3.32 to 3.40, showing that conditional expert routing improves the adaptive-depth configuration without changing its mean refinement depth substantially.

Adding EIS shifts the adaptive model toward substantially shallower execution. The mean refinement depth decreases from 2.93 to 2.15, while the average computation decreases from 26.71 to 20.73~GMACs/s. Meanwhile, WB-PESQ changes from 3.40 to 3.37. These results indicate that direct supervision of intermediate outputs enables the ADC to select earlier exits with a limited additional quality loss.

\begin{figure}
    \centering
    \includegraphics[width=1\linewidth]{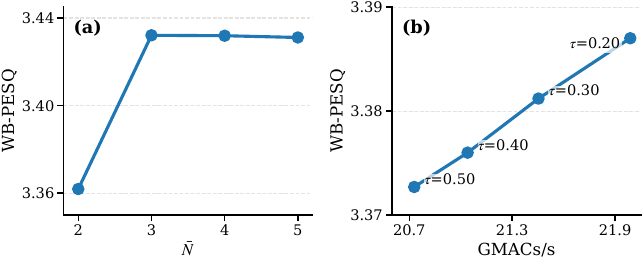}
    \caption{Analysis of refinement depth and adaptive computation in ADER.
    (a) WB-PESQ under different forced depths $\bar N$.
    (b) Computation--quality trade-off under different ADC continuation threshold $\tau$.}
    \label{fig:analysis}
\end{figure}

\subsection{Refinement Depth and Adaptive Computation}
\label{sec:mechanism}

Figure~\ref{fig:analysis}(a) evaluates the EIS-trained model with ADC disabled and a fixed refinement depth for all test utterances. WB-PESQ increases from 3.36 at $\bar N=2$ to 3.43 at $\bar N=3$, while further refinement produces little additional improvement. This result indicates that the perceptual benefit of progressive refinement largely saturates after the third iteration, supporting the use of input dependent termination rather than a fixed maximum depth for every utterance.

Figure~\ref{fig:analysis}(b) shows the computation--quality trade-off obtained by varying the ADC continuation threshold without changing the model parameters. Increasing $\tau$ from 0.20 to 0.50 reduces the mean refinement depth from 2.33 to 2.15 and the average computation from 21.99 to 20.73~GMACs/s, while WB-PESQ decreases from 3.387 to 3.372. Thus, the continuation threshold provides a direct control over the average inference cost through hard input-dependent termination.

\section{Conclusion}
\label{sec:conclusion}

We presented ADER for efficient SE with parameter-shared progressive refinement and input-dependent execution. By combining ADC, CER, and EIS, ADER supports hard early termination while retaining a compact shared refinement backbone. Experiments on VCTK-DEMAND show that ADER substantially reduces model size and average computation, with the continuation threshold providing a controllable trade-off between computational cost and enhancement quality. Forced-depth evaluation further shows limited WB-PESQ gains beyond the third refinement iteration, while the component analysis demonstrates the complementary roles of CER and EIS under adaptive-depth inference. These results show that adaptive block reuse can reduce redundant computation in SE. The current evaluation is limited to VCTK-DEMAND and an MP-SENet backbone. Future work will examine ADER on broader datasets and other SE architectures to assess its general applicability.


\bibliographystyle{IEEEbib}
\bibliography{strings,refs}

\end{document}